# Swelling of acetylated wood in organic liquids

Eiichi OBATAYA* and Joseph GRIL**
* Institute of Wood Technology, Akita Prefectural University 016-0876 Akita Japan
** Laboratoire de Mécanique et Génie Civil, Université Montpellier 2

**Abstract**

To investigate the affinity of acetylated wood for organic liquids, Yezo spruce wood specimens were acetylated with acetic anhydride, and their swelling in various liquids were compared to those of untreated specimens. The acetylated wood was rapidly and remarkably swollen in aprotic organic liquids such as benzene and toluene in which the untreated wood was swollen only slightly and/or very slowly. On the other hand, the swelling of wood in water, ethylene glycol and alcohols remained unchanged or decreased by the acetylation. Consequently the maximum volume of wood swollen in organic liquids was always larger than that in water. The effect of acetylation on the maximum swollen volume of wood was greater in liquids having smaller solubility parameters. The easier penetration of aprotic organic liquids into the acetylated wood was considered to be due to the scission of hydrogen bonds among the amorphous wood constituents by the substitution of hydroxyl groups with hydrophobic acetyl groups.

## Introduction

Many researchers have dealt with the swelling of wood in various organic liquids for comprehensive understanding of wood-liquid interaction. Neyer and Hossfeld[1] found that the hydrogen bonding properties of organic liquids affected the swelling of wood. The important role of hydrogen bonding was also suggested by Horiike and Kato[2] who measured the heat of wetting of wood in various organic liquids. The following investigations by Ishimaru and Adachi[3] indicated that the wood swelling in liquids having only a proton acceptor was larger than that in liquids having both a proton acceptor and donor, and that larger molar volume of liquid gave less swelling of wood when compared at the same proton accepting power. Meanwhile Mantanis *et al.*[4,5] performed a precise measurement on the swelling rate of wood in various liquids and they concluded that the apparent activation energy in the swelling process well corresponded to that for the scission of internal hydrogen bonds among wood constituents. In recent thermodynamic approaches by Morisato *et al.*[6-8] they discussed the cohesive interaction within the bulk liquids in the phased adsorption processes of wood. Although the correlations between wood swelling and liquid properties are not always clear, it is generally accepted that the swelling of wood involves competitive processes of the adsorption by hydrogen bonding and the scission of internal hydrogen bonds between the amorphous molecules of wood constituents, therefore it depends on the hydrogen bonding properties of liquids.

On the other hand, various chemical modifications have so far been proposed to improve the practical performances of wood. Among these, the acetylation is a typical method involving the chemical reaction between the wood constituents and chemicals. By the acetylation, the hydroxyl groups in the amorphous region of wood cell wall are substituted by acetyl groups. Consequently, the acetylated wood shows excellent dimensional stability and durability for the bulky and hydrophobic nature of acetyl groups introduced.[9]

It should be remembered that the acetylation is not only applied for the wood modification but also widely used to improve the compatibility of hydrophilic materials with hydrophobic substances. In fact, we have recently found that a bulky hydrophobic glucose pentaacetate (GPA: $M_W$=390) could be easily introduced in to the acetylated wood cell wall, whereas it can hardly penetrate into the untreated wood cell wall.[10,11] This fact suggests that the acetylated wood can also be swollen in organic liquids in which the untreated wood shows no or very small swelling. If so, various hydrophobic chemicals so far untested might be introduced into the wood cell wall by using an appropriate organic solvent, to expand the possible application of the acetylated wood. However, little information is available for the swelling of acetylated wood in organic liquids at this time. In this paper, we compare the swelling of untreated and acetylated woods in various organic liquids to exhibit the drastic changes due to the acetylation.

## Materials and method

Two ezomatsu (*Picea yezoensis*) wood logs (I and II) were sliced into two sets of specimens with a size of 1cm (L, longitudinal direction) by 3cm (R, radial direction) by 3cm (T, tangential direction), respectively. The specimens were previously washed in water and acetone using Soxhlet extractor for 8hours to remove the extractives. These specimens were then absolutely dried *in vacuo* at room temperature, and their weight and dimension were measured. The absolutely dried specimens from the log I were soaked in acetic anhydride overnight, and heated in a separable flask at 120°C for 8hrs. The specimens were immediately cooled at room temperature, and then washed in flowing water for a week to remove the chemical reagents and acetic acid remaining there. The specimens were finally dried completely *in vacuo* at room temperature and their weights and dimensions were measured again. Next the specimens were soaked under vacuum in various liquids listed in Table 1. All liquids other than Wt and Eg were analytical grade for gas or liquid chromatography. The liquids and specimens were put into grass bottles being tightly closed with polytetrafluoroethylene packing. The bottles were then kept at 20°C over 240days, and the dimension of specimens was measured intermittently. Six specimens were used for each testing condition.

The another set of specimens from the log II was separated into 4 groups. One group remained unacetylated and the other groups were acetylated at 120°C for 10min, 1hr, and 8hr by the same manner described above. The specimens were then washed in water and dried absolutely *in vacuo*. Those untreated and acetylated specimens were soaked in Be, To, Ac, Me, Et, Pr, Bu and Wt at 20°C for 60days and their dimensions were measured. Three specimens were used for each testing condition.

**Table 1.** Characteristics of liquids used.

| Ab. | Liquids | $M_W$ | $M_V$ (cm³/mol) | SP [(cal/cm³)^{1/2}] |
|---|---|---|---|---|
| Be | Benzene ($C_6H_6$) | 78.1 | 89.4 | 9.2 |
| To | Toluene ($C_6H_5CH_3$) | 92.1 | 106.4 | 8.9 |
| Am | Methyl acetate ($CH_3COOCH_3$) | 74.1 | 79.7 | 9.6 |
| Ae | Ethyl acetate ($CH_3COOC_2H_5$) | 88.1 | 98.5 | 9.1 |
| Ac | Acetone ($CH_3\text{-}CO\text{-}CH_3$) | 58.1 | 74.0 | 9.9 |
| Di | 1,4-Dioxane ($C_4H_8O_2$) | 88.1 | 85.7 | 10.0 |
| Me | Methanol ($CH_3OH$) | 32.0 | 40.7 | 14.5 |
| Et | Ethanol ($C_2H_5OH$) | 46.0 | 58.5 | 12.7 |
| Pr | 2-Propanol (($CH_3)_2CHOH$) | 60.1 | 75.0 | 11.5 |
| Bu | 1-Butanol ($CH_3CH_2CH_2CH_2OH$) | 74.1 | 91.8 | 11.4 |
| Eg | Ethylene glycol ($HOCH_2CH_2OH$) | 62.1 | 55.8 | 14.6 |
| Df | N,N-Dimethylformamide ($HCON(CH_3)_2$) | 73.1 | 77.0 | 12.1 |
| Ds | Dimethyl sulphoxide (($CH_3)_2SO$) | 78.1 | 70.9 | 12.0 |
| Wt | Water ($H_2O$) | 18.0 | 18.0 | 23.4 |

$M_W$, Molecular weight; $M_V$, molar volume; SP, solubility parameter.[12]

**Results and discussion**

The basic characteristics of wood specimens tested are listed in Table 2. The swelling of untreated specimen in water was evaluated by

$$\Delta V_W = 100(V_W - V_0)/V_0,$$

where the $V_0$ and $V_W$ are the volume of untreated wood in the absolutely dry condition and that in water, respectively. Although the different logs gave different $\Delta V_W$ values, the variation within the same log was small enough to assume the homogeneity of specimens in their swelling property.

**Table 2.** Density ($\rho_U$) and volumetric swelling in water ($\Delta V_W$) for the untreated wood specimens, and density ($\rho_A$), weight percent gain (WPG) and swelling in volume ($\Delta V_A$) for the acetylated wood specimens.

| Log | Treat | t (hr) | $\rho_U$ (g/cm³) | $\Delta V_W$ (%) | $\rho_A$ (g/cm³) | WPG (%) | $\Delta V_A$ (%) |
|---|---|---|---|---|---|---|---|
| I | U | 0 | **0.407** 0.006 | 17.6 0.3 | | | |
| | A | 12 | **0.404** 0.008 | 17.6 0.4 | 0.471 0.010 | 26.5 0.4 | 8.1 0.4 |
| II | U | 0 | **0.374** 0.008 | 19.2 0.3 | | | |
| | | 0.17 | **0.377** 0.008 | 19.4 0.2 | 0.385 0.016 | 5.9 1.9 | 0.6 1.0 |
| | A | 1 | | | 0.396 0.017 | 15.4 0.9 | 5.4 0.2 |
| | | 12 | | | 0.411 0.012 | 26.4 1.0 | 7.6 0.4 |

t, Treating duration; values in the upper and lower columns indicate the average and standard deviation.

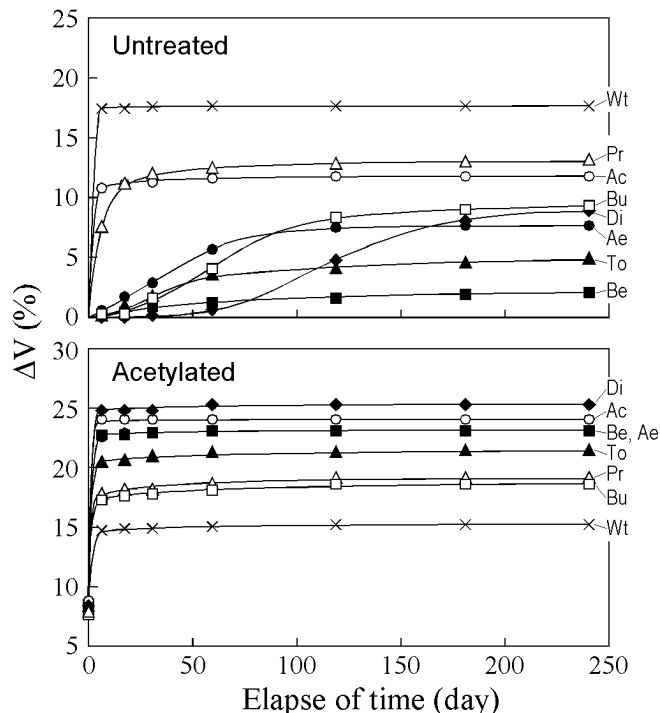

**Figure 1.** Swelling in volume of untreated (upper) and acetylated (lower) wood specimens soaked in various organic liquids plotted against the soaking duration. Abbreviations indicates the liquids listed in Table 1. The $\Delta V$ is based on the volume of wood specimen in its untreated and absolutely dry state.

The degree of acetylation was evaluated by weight percent gain (WPG) and swelling in volume ($\Delta V_A$). The $\Delta V_A$ was defined by

$$\Delta V_A = 100(V_A - V_0)/V_0,$$

where the $V_A$ is the absolutely dry volume of the acetylated wood specimen. The variations in WPG and $\Delta V_A$ were very small, except for those by short treatment (10min). It was thought that the slighter weight and volume gains due to the short treatment were compensated with the thermal degradation of wood components or the removal of extractives in the early stage of treatment, to result in wider variation in WPG and $\Delta V_A$.

Figure 1 shows the swelling of untreated and acetylated wood specimens soaked in various liquids over 240 days. The volumetric swelling of wood was evaluated by

$$\Delta V = 100(V_S - V_0)/V_0,$$

where the $V_S$ is the volume of wood swollen in liquids. The volume of untreated wood in Wt reached its maximum within a day, but it showed slight and/or very slow increase in aprotic organic liquids such as Be, To, Ae, Di, and also in protic alcohols, Bu and Pr, having large molar volume. On the other hand, the acetylated woods soaked in organic liquids were swollen very rapidly, and their maximum swellings were always greater than that in Wt.

In Figure 2, the swelling of wood by 240days of soaking are plotted against the solubility parameter (SP) of liquids. According to Ishimaru and Robertson, organic liquids can be classified into two groups in respect of their effects on wood swelling[3] and decreasing strength of paper[13]: group I, liquids having only a proton acceptor; group II, liquids having both a proton acceptor and donor. The former gives greater swelling than the latter at the same SP, proton accepting power (PAP) or PAP divided by molar volume. The plots of untreated wood followed the trend in earlier reports *i.e.* the ΔV in group I liquids were always larger than those in group II liquids at the same SP, and clear correlation between ΔV and SP was obtained for each group. Except for polar aprotic liquids such as Ds, liquids having smaller SP resulted in smaller swelling of untreated wood.

On the other hand, the swollen volume of wood increased drastically after the acetylation, except for that in Eg and Wt where the acetylated wood showed slightly less swelling. Such changes were greater in liquid having smaller SP value, consequently, the swollen volume of acetylated wood was less dependent on the SP value. The drastic change in ΔV due to the acetylation can be interpreted by the following two factors: easier adsorption of non- or low-polar liquid molecules onto the wood where the low-polar acetyl groups are introduced; less energy to break the internal hydrogen bonds among the wood constituents by the substitution of hydroxyl groups with the acetyl groups. Probably both of them affect the swelling of wood though, but detailed thermodynamic approach must be needed for further discussion.

have been obtained by Ishimaru and Maruta,[14] and explained by the restriction of radial swelling by the heterogeneous cell wall structure such as the different reorientation of cellulose microfibrils in radial and tangential cell walls.

The swelling of wood due to the acetylation was quite anisotropic (ΔT=6.6%, ΔR=1.5%). Such greater anisotropy was kept in various liquids, and the radial swelling of the acetylated wood seemed to be less restricted than that of the untreated wood at higher swelling level. Among major wood constituents, lignin and hemicelluloses substituted for readily at small WPGs, whereas the cellulose in wood is less reactive on the acetylation.[15-16] Therefore, it was speculated that the location of lignin rich part, such as middle layer, is responsible for the anisotropic swelling of wood due to the acetylation and the following soaking in organic liquids. This speculation should be ascertained by microscopic observation on the acetylated wood in various swollen states.

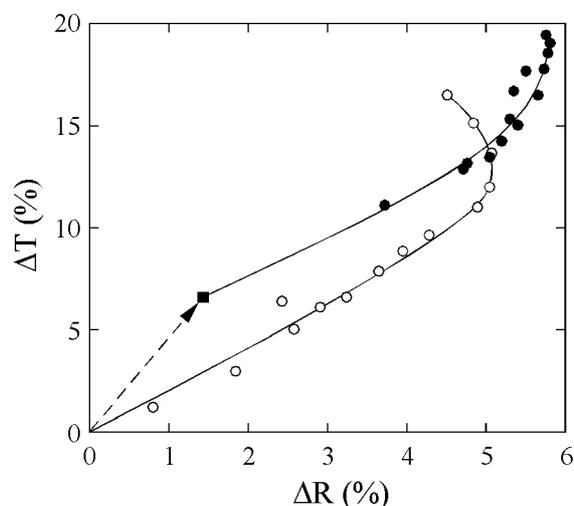

**Figure 3.** Relationship between the tangential swelling (ΔT) and the radial swelling (ΔR) of untreated and acetylated wood specimens soaked in various liquids. Open circles, untreated wood swollen in liquids; filled circles, acetylated wood swollen in liquids; filled square, absolutely dry acetylated wood.

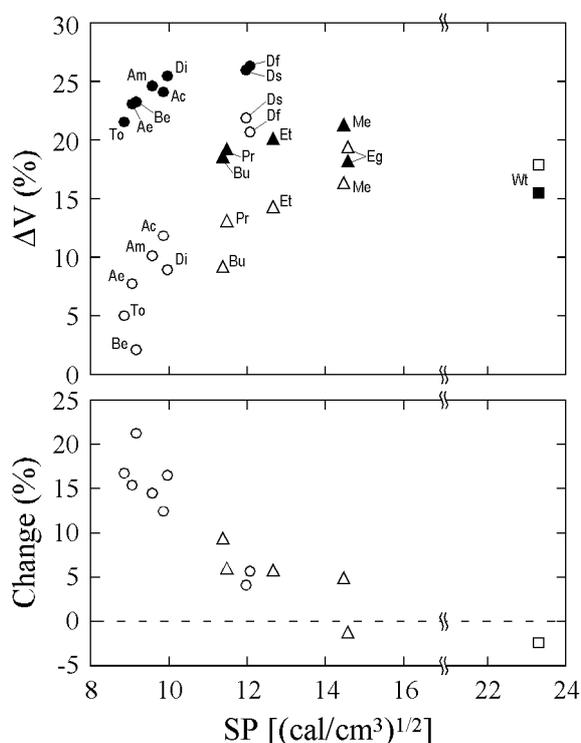

**Figure 2.** Effects of solubility parameter (SP) of liquids on the swelling in volume (ΔV) of wood specimens (upper), and changes in ΔV due to the acetylation (lower). Circles, aprotic organic liquids; triangles, protic organic liquids (alcohol and glycol); squares, water; open plots, untreated wood; filled plots, acetylated wood. For abbreviations, see Table 1.

Figure 3 exhibits the radial and tangential swellings of untreated and acetylated wood specimens in various liquids. Both the ΔR and ΔT values were calculated based on the dimension of wood specimens in their untreated and absolutely dried state. The tangential swelling of untreated wood was almost proportional to its radial swelling up to 10% (ΔR=5%), above which the radial swelling was restricted. Such a turning point corresponded to the swelling of wood in water (ΔT=12.0%, ΔR=5.1%). Similar results

Figure 4 shows the effects of WPG on the ΔV of wood. The 60days of soaking was not long enough to achieve the maximum swelling, especially for the untreated wood soaked in organic liquids, but enough to exhibit the drastic changes due to the acetylation. The swollen volume of wood in aprotic liquids increased with increasing WPG (Fig.4A), and the ΔV values were converged to a certain range from 20% to 25% irrespective of liquids. That range might be a limit determined by the fiber-reinforced structure of the wood cell wall. On the other hand, the effects of WPG was relatively small when wood was soaked in protic liquids such as alcohol, glycol and water (Fig.4B).

It should be remembered that the ΔV value for the acetylated wood includes the initial swelling due to the acetylation. Thus here we define another parameter by

$$\Delta V_S = 100(V_S - V_A)/V_0$$

to evaluate the swelling of wood due to the soaking in liquids. The ΔV$_S$ values are plotted against the WPG in Figure 5. The difference between the aprotic and protic liquids became more clear. The ΔV$_S$ values in aprotic liquids increased with increasing WPG whereas those in protic liquids showed reversed trend. The excellent dimensional stability of acetylated wood is generally explained by the occupation of intermolecular space in the amorphous region of wood cell wall, so called "bulking effect," but such an effect may be valid only for the liquids having greater hydrogen bonding powers.

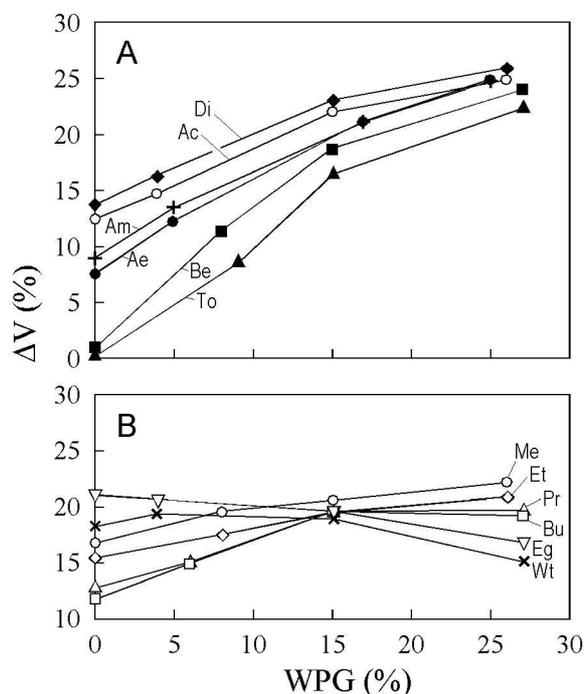

**Figure 4.** Effects of weight percent gain (WPG) on the swelling in volume (ΔV) of wood specimens soaked in aprotic (A) and protic (B) liquids for 60days. For abbreviations, see Table 1.

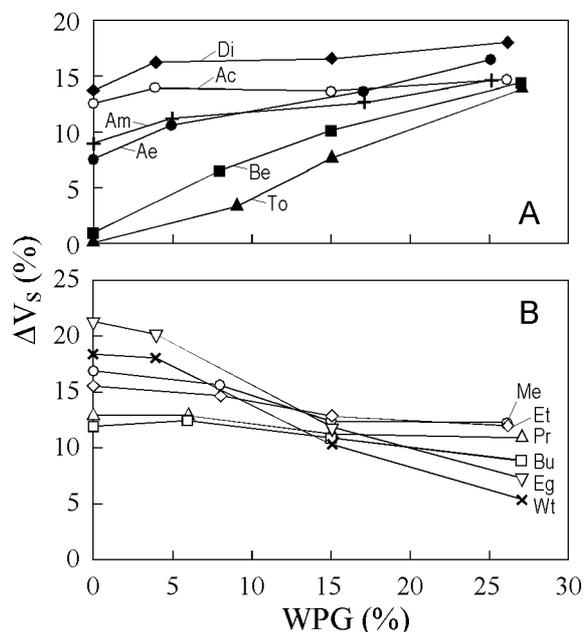

**Figure 5.** Swelling in volume of wood specimens due to the penetration of liquids ($\Delta V_S$) plotted against the weight percent gain (WPG) due to the acetylation. For abbreviations, see Table 1.

**Conclusion**

The swellings of acetylated wood were compared to those of the untreated wood in various organic liquids. The acetylated wood was swollen rapidly and remarkably in organic liquids in which the untreated wood was swollen only slowly and slightly. Clear correlation was obtained between the effect of acetylation on the swollen volume of wood and the solubility parameter of liquids. The acetylation resulted in greater anisotropy in its transverse swelling which was kept in various liquids. The swelling of wood in aprotic liquids increased with increasing the degree of acetylation while that in protic liquids showed reversed trend.